\documentclass{aa}

\usepackage{graphicx}
\usepackage{txfonts}
\usepackage{natbib}

\begin{document}

\title{Orbits and Masses in the multiple system LHS 1070%
\thanks{Based on observations collected at the European Southern
  Observatory, Chile, proposals number 60.A-9026, 66.C-0219,
  67.C-0354, 68.C-0539, 70.C-0476, 072.C-0022, 074.C-0637,
  078.C-0386, 380.C-0179, 382.C-0324, and 382.C-0329.}}

\author{R. K\"ohler\inst{1,2}
	\and
	T. Ratzka\inst{3}
	\and
	Ch. Leinert\inst{1}
}

\institute{%
	Max-Planck-Institut f\"ur Astronomie, K\"onigstuhl 17,
	69117 Heidelberg, Germany,
        \email{koehler@mpia.de}
\and
	Landessternwarte, Zentrum f\"ur Astronomie der Universit\"at Heidelberg,
	K\"onigstuhl, 69117 Heidelberg, Germany,
        \email{r.koehler@lsw.uni-heidelberg.de}
\and
	Universit\"ats-Sternwarte M\"unchen,
	Ludwig-Maximilians-Universit\"at,
	Scheinerstr. 1, 81679 M\"unchen, Germany
}

\date{Received 22 December 2011; accepted 15 March 2012}

\abstract{}{We present a study of the orbits of the triple system
  LHS\,1070, with the aim to determine individual masses of its
  components.}
{Sixteen new relative astrometric positions of the three components in
  the K band were obtained with NACO at the VLT, Omega CASS at the
  3.5\,m telescope on Calar Alto, and other high-spatial-resolution
  instruments.  We combine them with data from the literature and fit
  orbit models to the dataset.
  We derive an improved fit for the orbit of LHS\,1070\,B and C around
  each other, and an estimate for the orbit of B and C around A.}
{The orbits are nearly coplanar, with a misalignment angle of less
  than $10^\circ$.
  The masses of the three components are
  $M_A=0.13\ldots0.16\rm\,M_\odot$, $M_B = 0.077\pm0.005\rm\,M_\odot$,
  and $M_C = 0.071\pm0.004\rm\,M_\odot$.  Therefore, LHS\,1070\,C is
  certainly, and LHS\,1070\,B probably a brown dwarf.  Comparison with
  theoretical isochrones shows that LHS\,1070\,A is either fainter or
  more massive than expected.  One possible explanation would be that
  it is a binary.  However, the close companion reported previously
  could not be confirmed.
}{}


\maketitle


\section{Introduction}

\object{LHS 1070}
(other common names are \object{GJ 2005}, \object{LP 881-64},
\object{2MASS J00244419-2708242})
is a nearby high-proper-motion star located in the south galactic
pole region.
It was observed by \citet{leinert1994} as part of their near-infrared
speckle survey for duplicity of nearby southern M dwarves.
The group detected in 1993 two companions located 1.1" and 1.3" north
of the primary by using the SHARP camera mounted at the NTT
(La Silla, Chile) and applying Speckle imaging. In a consecutive
measurement the companions exhibited an orbital motion around each
other and around the primary while following the proper motion of the
system. A fourth component D was identified by \citet{henry1999} with
the Fine Guidance Sensors onboard the Hubble Space Telescope (HST).  A
separation of only about 50\,mas from component A was reported. Speckle
interferometric and adaptive optics measurements, however, were
not able to confirm LHS\,1070\,D \citep{leinert01, seifahrt2008}.
This might be due to the small separation, but it is also possible
that the detection is spurious (T.~Henry, priv.\ comm.).

When comparing the colours of the components, the companions appear
redder than the primary \citep{leinert1994}. First estimates already
identified the two companions as stars close to the hydrogen burning
limit that divides brown dwarves from main-sequence stars. This mass
range is characterised by a strong decrease of the effective
temperature and the onset of dust formation in the atmospheres.
Photometric and spectroscopic measurements with the HST in the visual
indeed were represented by model atmospheres containing dust
\citep{leinert2000}. Also the low masses could be confirmed.
The spectral classifications of the companions are M8.5V and
M9-9.5V. The primary has an earlier spectral type of M5.5-6V.

\citet{reiners2007} found for the components B and C a similarily high
rotational velocity of $v\sin i = 16$\,km/s, which is twice that of
the primary. The authors conclude that the temperature-dependent
magnetic braking was acting for 1\,Gyr on all three components. The
main difference between the otherwise very similar components B and C
is the higher activity of component B. This might be related to the
higher temperature or the higher magnetic flux. In the HST spectra
H$\alpha$ emission was found towards LHS\,1070\,A and B, but not
towards component C \citep{leinert2000}. Recently, photometric
activity of the two more massive stars was identified
\citep{almeida2011}.  Component B showed a brightness increase in the
visual that is among the largest ever observed in a flare star.

The very low mass companions of LHS\,1070 provide the possibility to
refine the models of (pre-)main-sequence stars close to the transition
region to brown dwarves. It is thus very important to derive the exact
masses of these stars. The small separation of the components B and C,
i.e.\ their short orbital period is very well suited to allow even for
this low mass stars a mass determination by fitting the orbital
elements.
The dynamical mass has the advantage of being independent from
theoretical models. It is thus a precious probe to test evolutionary
models.

%
\begin{table*}[ht]
\caption{Astrometric measurements of LHS 1070 B-C}
\label{ObsTab}
\begin{center}
\begin{tabular}{ll@{${}\pm{}$}lr@{${}\pm{}$}ll@{${}\pm{}$}lr@{${}\pm{}$}ll}
\noalign{\vskip1pt\hrule\vskip1pt}
 Date (UT) 	& \multicolumn{4}{c}{A-B}	& \multicolumn{4}{c}{B-C} & Reference, Instrument, Proposal-No.\\
	& \multicolumn{2}{c}{$d$ [mas]}	& \multicolumn{2}{c}{PA~$[^\circ]$}
 						& \multicolumn{2}{c}{$d$ [mas]} & \multicolumn{2}{c}{PA~$[^\circ]$} &\\
\noalign{\vskip1pt\hrule\vskip1pt}
 1993 Jul 29	    	& $1072$  &$10$ & $-8.6$&$0.3$ 	& $266. $&$ 5.$ & $328.5$&$0.7 $ & \cite{leinert01}\\
 1994 May  1	    	& $1085$  & $4$ & $-7.7$&$0.1$ 	& $341. $&$ 5.$ & $346.9$&$0.6 $ & \cite{leinert01}\\
 1994 Sep 15	    	& $1092$  & $9$ & $-7.2$&$0.4$ 	& $375. $&$ 4.$ & $352.7$&$0.61$ & \cite{leinert01}\\
 1994 Sep 24	    	& $1095$  &$12$ & $-6.8$&$0.42$	& $382. $&$ 5.$ & $353.9$&$0.67$ & \cite{leinert01}\\
 1995 Jan 17	    	& $1094$  & $6$ & $-6.2$&$0.3$ 	& $400. $&$ 8.$ & $356.1$&$0.4 $ & \cite{leinert01}\\
 1995 Jul  9	    	& $1119$  &$11$ & $-6.4$&$0.1$ 	& $439. $&$ 4.$ & $  3.2$&$0.1 $ & \cite{leinert01}\\
 1995 Jul 14	    	& $1102$  & $3$ & $-6.1$&$0.1$ 	& $436. $&$ 1.$ & $  3.7$&$0.2 $ & \cite{leinert01}\\
 1996 Jan 16	    	& $1124$  & $2$ & $-5.6$&$0.3$ 	& $459. $&$ 3.$ & $  9.0$&$0.3 $ & \cite{leinert01}\\
 1996 Aug 22	    	& $1161$  & $5$ & $-5.0$&$0.1$ 	& $465. $&$ 3.$ & $ 14.9$&$0.1 $ & \cite{leinert01}\\
 1996 Sep 27	    	& $1157$  & $8$ & $-4.7$&$0.1$ 	& $468. $&$ 4.$ & $ 15.8$&$0.1 $ & \cite{leinert01}\\
 1997 Jul 15	    	& $1235$  &$14$ & $-3.5$&$0.23$	& $458. $&$ 6.$ & $ 23.4$&$0.6 $ & \cite{leinert01}\\
 1997 Aug 25	    	& $1243$  & $7$ & $-3.5$&$0.23$	& $450. $&$15.$ & $ 25.4$&$1.0 $ & \cite{leinert01}\\
 1997 Nov 17	    	& $1223$  & $5$ & $-2.9$&$0.1$ 	& $439. $&$ 8.$ & $ 26.5$&$0.45$ & \cite{leinert01}\\
 1998 Jan  2	    	& $1260$  & $3$ & $-2.5$&$0.14$	& $432. $&$ 3.$ & $ 28.0$&$0.45$ & \cite{leinert01}\\
 1998 May  7	    	& $1281$  & $7$ & $-1.9$&$0.1$ 	& $408. $&$ 8.$ & $ 32.3$&$0.73$ & \cite{leinert01}\\
 1998 Oct 10	    	& $1332$  & $8$ & $-1.0$&$0.1$ 	& $377. $&$19.$ & $ 38.2$&$0.5 $ & \cite{leinert01}\\
 1999 Jun 18	    	& $1404$  & $3$ & $0.44$&$0.1$ 	& $318. $&$ 2.$ & $ 49.4$&$0.4 $ & \cite{leinert01}\\
 1999 Aug  3	    	& $1407$  & $7$ & $1.02$&$0.28$	& $303. $&$ 7.$ & $ 52.1$&$0.3 $ & \cite{leinert01}\\
 1999 Sep  1	    	& $1414$  & $4$ & $1.23$&$0.17$	& $292. $&$ 8.$ & $ 54.6$&$1.0 $ & \cite{leinert01}\\
 1999 Nov 23	    	& $1437$  & $4$ & $1.64$&$0.11$	& $279. $&$ 5.$ & $ 60.1$&$0.9 $ & \cite{leinert01}\\
 2000 Jun  4	    	& $1487$  & $9$ & $2.5$ &$0.2$ 	& $240. $&$ 2.$ & $ 76.0$&$0.4 $ & \cite{leinert01}\\
 2000 Jun 20	    	& $1518$  & $3$ & $3.6$ &$0.3$	& $237. $&$ 3.$ & $ 80.8$&$0.7 $ & \cite{leinert01}\\
 2000 Oct 31 -- Nov 1 	& $1516$  &$1.4$&  $4.4$&$0.1$	& $218.1$&$1.9$ & $ 88.9$&$0.6 $ & OCASS\\ 
 2001 Jan  9	    	& $1546.4$&$1.1$&  $4.7$&$0.1$	& $219.0$&$9.0$ & $100.4$&$2.3 $ & ADONIS, 66.C-0219	\\
 2001 June 29 -- July 6	& $1600.7$&$5.5$&  $5.4$&$0.1$	& $222.7$&$1.0$ & $120.3$&$0.2 $ & SHARP,  67.C-0354	\\
 2001 Dec  6		& $1631.2$&$2.5$&  $6.8$&$0.2$	& $244.9$&$0.4$ & $136.1$&$0.2 $ & NACO, 60.A-9026 (commissioning) \\
 2001 Dec 11		& $1633.0$&$7.7$&  $7.1$&$0.4$	& $246.7$&$3.8$ & $136.8$&$0.9 $ & ADONIS, 68.C-0539	\\
 2002 Dec 16		& $1712.1$&$3.4$&  $9.8$&$0.3$	& $327.7$&$0.9$ & $162.8$&$0.3 $ & NACO, 70.C-0476	\\
 2003 Dec 12		& $1772.3$&$4.9$& $12.7$&$0.1$	& $406.3$&$1.1$ & $178.2$&$0.1 $ & NACO, 072.C-0022	\\
 2004 Dec 11		& $1796.3$&$2.2$& $15.9$&$0.1$	& $450.0$&$1.4$ & $189.7$&$0.2 $ & NACO, 074.C-0637	\\
 2005 Sep 25		& $1792. $&$19.$& $17.9$&$0.2$	& $450. $&$11.$ & $198.7$&$0.9 $ & OCASS		\\
 2006 Sep  1 -- 2	& $1759.2$&$4.6$& $20.5$&$0.2$	& $403.1$&$6.0$ & $208.1$&$0.9 $ & OCASS		\\
 2006 Oct 30		& $1742.9$&$2.9$& $20.9$&$0.5$	& $399.7$&$1.0$ & $209.6$&$0.5 $ & NACO, 078.C-0386	\\
 2006 Nov 28		& $1744.7$&$3.8$& $21.3$&$0.2$	& $396.1$&$2.0$ & $211.0$&$0.3 $ & OCASS		\\
 2007 Sep 16		& $1680.2$&$1.5$& $23.6$&$0.2$	& $328.7$&$0.4$ & $223.5$&$0.2 $ & NACO, 380.C-0179	\\
 2008 Feb  1		& $1646.4$&$1.4$& $24.8$&$0.2$	& $295.4$&$0.4$ & $231.5$&$0.2 $ & NACO, 380.C-0179	\\
 2008 Oct 17,18		& $1576.6$&$0.7$& $27.3$&$0.1$	& $240.7$&$0.3$ & $252.6$&$0.13$ & NACO, 382.C-0324/0329\\
 2008 Nov  6		& $1571.3$&$0.7$& $27.4$&$0.1$  & $237.6$&$0.3$ & $254.6$&$0.12$ & NACO, 382.C-0324	\\
\noalign{\vskip1pt\hrule\vskip1pt}
\end{tabular}
\end{center}
\end{table*}


The main uncertainty for the mass determination is the distance to the
object. The trigonometric parallax of LHS\,1070 derived by
\cite{altena1995} is $135.3\pm 12.1$\,mas. This value was later
refined to $129.47\pm 2.48$\,mas, placing the system at a distance of
$7.72\pm 0.15$\,pc \citep{costa2005}. The proper motion of LHS\,1070 is
{$653.7\pm0.3$\,mas/yr}, corresponding to 23.9\,km/s, and directed
along the postion angle $348.3^{\circ}\pm0.44^{\circ}$. In addition,
radial velocity measurements found that LHS\,1070 is approaching with
36.4 km/s to the Sun \citep{basri1995}. Due to its kinematic
properties, LHS\,1070 can be associated with the old disk population.

A first successful fit of the close orbit of component C around
component B was presented by \citet{leinert01}. Based on Speckle
interferometric measurements a semi-major axis of $a=446\pm29$\,mas
and a period of $P=16.1\pm1.4$\,yr were derived. When taking the refined
distance of \cite{costa2005} into account, the corresponding dynamical
mass is $0.157\pm0.042\,M_{\odot}$. A consistent mass of
$0.157\pm0.009\,M_{\odot}$ was found by \cite{seifahrt2008}, who
derived a wider ($a=461.9\pm0.7$\,mas) orbit, but with
$P=17.0$\,yr. Due to the better coverage the authors presented also a
first reliable fit of the wide orbit of the components B and C around
the primary. The dynamical mass of the whole system was determined as
$0.272\pm 0.017\,M_{\odot}$. The orbits were found to be coplanar.

In this study, we present new measurements taken both with
Speckle interferometric techniques and adaptive optics in the K-band
(Section~2). The data is carefully selected and calibrated on a case
by case basis with the aim to reach the highest degree of consistency
possible. These new data are not only used to further refine the fits
of the close (Section~3) and the wide orbit (Sections~4). We utilise
the primary component as astrometric reference, which enables the
determination of the mass ratio of the components B and C. With this
knowledge the individual masses of all three components can be
derived. In Section~5 we discuss the implications of our findings
with respect to the stability of the system and its evolutionary
stage. We conclude in Section~6.


\begin{figure*}[t]
\centerline{%
  \includegraphics[width=0.25\hsize]{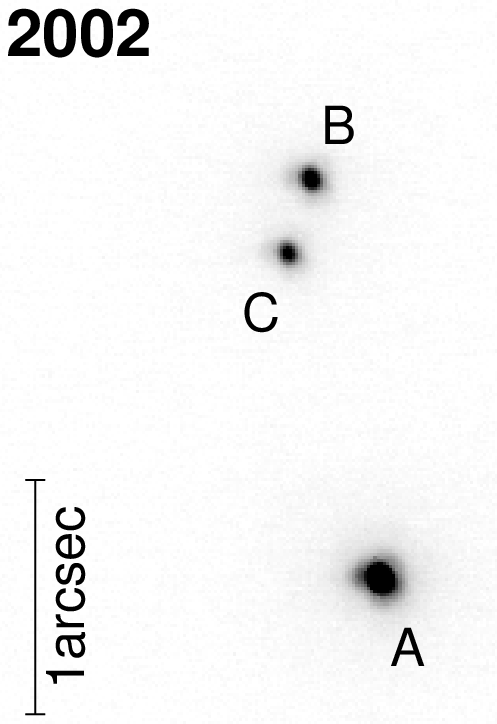}\hss
  \includegraphics[width=0.25\hsize]{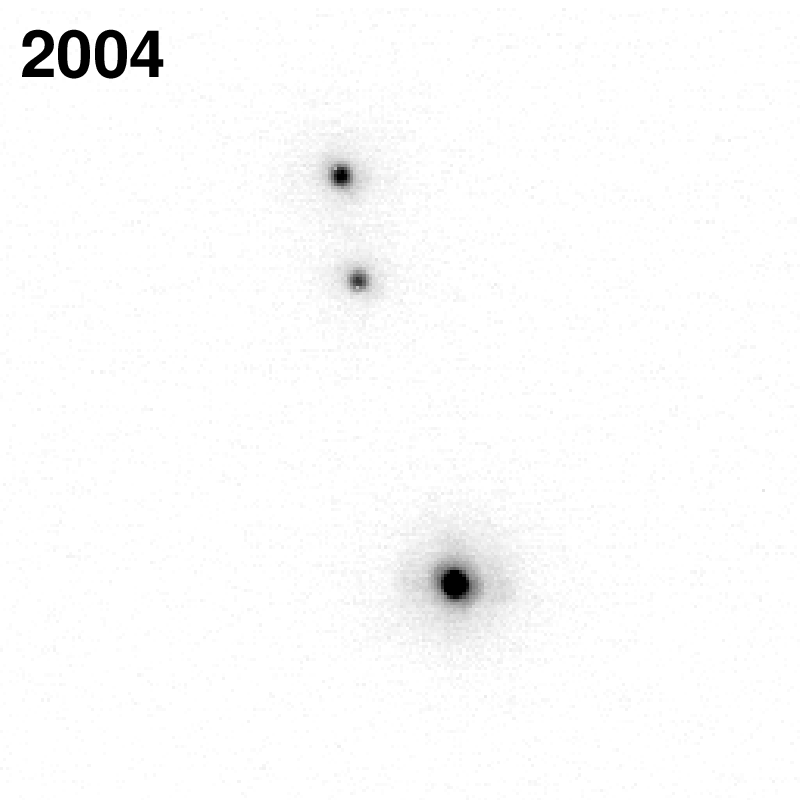}\hss
  \includegraphics[width=0.25\hsize]{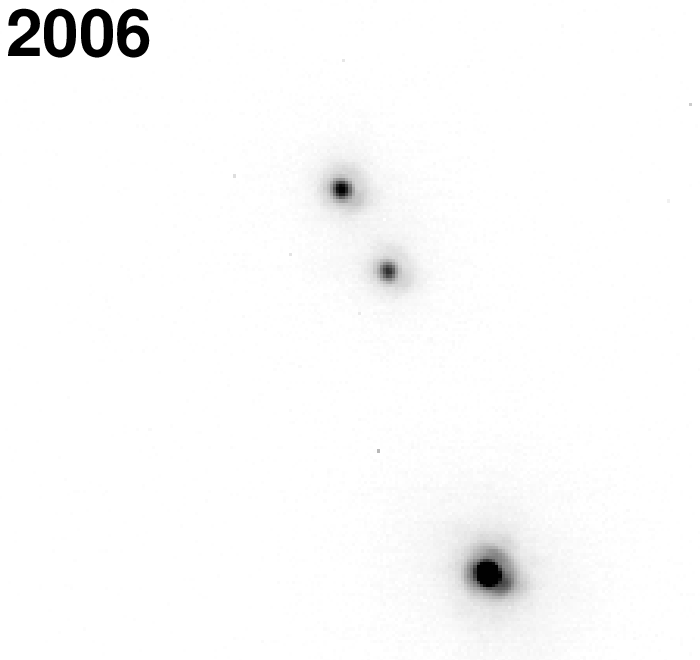}\hss
  \includegraphics[width=0.25\hsize]{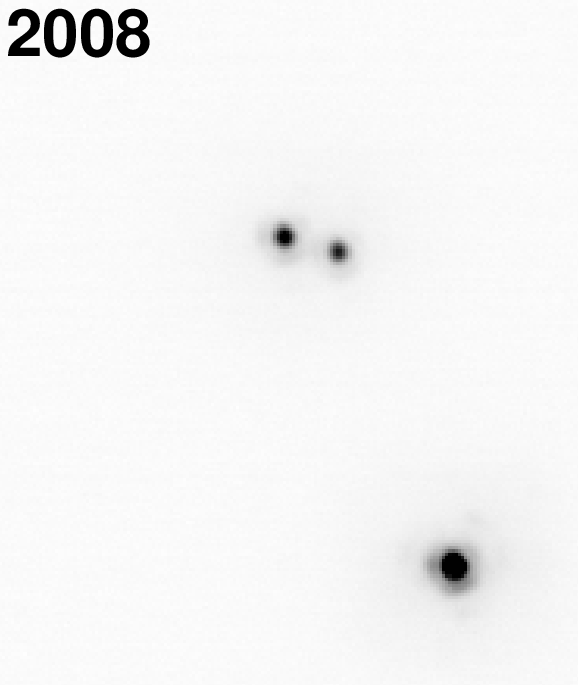}}
\caption{Images of LHS\,1070 obtained with NACO in December 2002,
  December 2004, October 2006, and November 2008.
  North is up, and East to the left.
  The separation between the two components B and C changes from about
  330\,mas in 2002 to about 240\,mas in 2008 (cf.\ Table~\ref{ObsTab}).
}
\label{NACOpic}
\end{figure*}


\section{Observations and data reduction}

Our group monitored LHS\,1070 since the discovery of its two
companions, using a number of different telescopes and instruments.
Table~\ref{ObsTab} gives a journal of observations and the measured
relative positions.  The following subsections describe the individual
instruments and data reduction procedures.

\subsection{ESO NTT / SHARP}

The SHARP~I camera (System for High Angular Resolution Pictures) of
the Max-Planck-Institut for Extraterrestrial Physics \citep{SHARP} was
already used for many of the observations presented in
\citet{leinert01}.
We used it again during an observing campaign in June and July 2001 at
the European Southern Observatory (ESO) 3.5\,m New Technology
Telescope (NTT) on La Silla, Chile.
LHS\,1070 was observed in the K-band at $2.2\rm\,\mu m$ in four
different nights.
SHARP is a camera for speckle interferometry, which means that one
observation consists of several hundred frames with short integration
times (0.5\,sec).
To reduce the data, we used our {\tt speckle} program
\citep{koehler2000}.
The same observing and data reduction strategy was already employed
for the data published in \citet{leinert01} and a number of other
multiplicity studies.

\subsection{ESO 3.6\,m / ADONIS / SHARP II}

In January and December 2001, LHS\,1070 was observed with the adaptive
optics (AO) system ADONIS \citep{ADONIS} and the SHARP~II camera
\citep{SHARPII} at the ESO 3.6\,m telescope on La Silla, Chile.
We used the K-band filter of this
instrument, which has a central wavelength of $2.177\rm\,\mu m$.
The observing strategy was ``AO-assisted speckle interferometry'',
meaning we took many frames with short integration times.
In January, we recorded 1000 frames with an integration time of
0.5\,sec each.  In December, we took 240 frames with a longer
integration time of 3\,sec each.
For the data reduction, we used the same programs and algorithms as
for the speckle data.

\subsection{Calar Alto 3.5\,m / Omega Cass}

On several occasions between 2000 and 2006, LHS\,1070 was observed
with the Omega Cass camera \citep{lenzen98} at the 3.5\,m-telescope on
Calar Alto, Spain.
The camera is equipped with a 1024$\times$1024-pixel detector, but we
used only a subarray of 128$\times$128 pixels to enable the fast
read-out mode required for Speckle-interferometry.
The observations were done in the K-band and with the {highest
resolution} optics of the instrument (pixel scale $\sim$95\,mas/pixel).
Our {\tt speckle} program was used again to reduce the data.

\subsection{ESO VLT / NACO}

The majority of new observations were taken with NAOS/CONICA (NACO for
short), the adaptive optics, near-infrared camera at the ESO Very
Large Telescope (VLT) on Cerro Paranal, Chile \citep{rousset03,
  lenzen03}.
LHS\,1070 was observed in the course of several programs (see
Table~\ref{ObsTab}).
To ensure a consistent data set, only imaging observations in the
$K_s$ photometric band were used for the orbit determination.

The NACO images were sky subtracted with a median sky image, and
bad pixels were replaced by the median of the closest good neighbors.
Finally, the images were visually inspected for any artifacts or
residuals.  Figure~\ref{NACOpic} shows an example of the results.
The {\tt Starfinder} program \citep{Diolaiti00} was used to measure
the positions of the stars.  The positions in several images taken
during one observation were averaged, and their standard deviation
used to estimate the errors.

\subsection{Plate scale and orientation}
\label{CalibSect}

For a heterogenous data set like this, it is crucial to calibrate the
absolute pixel scale and orientation of each observation.
To this end, we took images of fields in the Orion Trapezium
during each observing campaign, and reduced them in the same way as
the images of the science targets.  The measured positions of the
cluster stars were compared with the coordinates given in
\cite{mccaughrean94}.  The mean pixel scale and orientation were
computed from a global fit of all star positions.  The scatter of
values derived from star pairs were used to estimate the errors (see
sect.~\ref{ErrorSect}).\


The errors of the calibration are usually comparable to or larger than
the errors of the measured positions of the science target, indicating
the importance of a proper astrometric calibration.
For this reason, we decided to use in this work only data where images
of the Orion Trapezium were taken within a few days.
We have two observations of LHS\,1070 that were taken in June and
July, when Orion was not observable.  These observations are not used
here, in order to ensure a consistent calibration of all the data.
For the same reason, we do no use data from the literature that could
not be re-calibrated by our group.

{\subsection{Astrometric error estimates}
\label{ErrorSect}

Independent of instrument and telescope used, our observation resulted
in at least 4 images or speckle cubes for each epoch.  The final
result was obtained by averaging the relative positions measured in
individual images.  As estimate for the errors, we use the standard
deviation.

The error of the astrometric calibration was estimated in a similar
way.  We computed the pixel scale by dividing the separation of star
pairs in arcseconds by the measured separation in pixels.  Similarly,
the orientation is the difference of the position angle on the sky and
on the detector.  The standard deviation of the results for the many
star pairs in our calibration field was used as estimate for the
error.

An additional source of error might be the motion of the telescope
between the observation of the calibration field and the target.  This
should not cause much uncertainty for telescopes on a parallactic
mount (i.e.\ the Calar Alto 3.5\,m and the ESO 3.6\,m).  However,
telescopes on an alt-azimuth mount (ESO NTT and VLT) have to correct
for image rotation, which might introduce an error on the order of
$0.1^\circ$ (W.~Brandner, priv.\ comm.).  Additionally, the opening
and closing of the AO-loop, in connection with the active optics of
the primary mirror, might induce small changes in the image scale.
Our calibrations show that there are small changes in pixel scale and
orientation on the time scale of weeks or months, but -- to the best
of our knowledge -- there has been no study of short-term variations.

This unknown error source might explain why we seem to underestimate
our errors, which results in orbit fits with $\chi^2>1$
(cf.\ tables~\ref{OrbitBCtab} and \ref{OrbitABCtab}).
}


\section{The orbit of LHS\,1070 B-C}
\label{orbitBCsect}

We estimated the orbital parameters of the B-C pair by fitting orbit
models to all observations listed in Table~\ref{ObsTab}.  We followed
the procedure described in \citet{koehler2008}: a grid-search in
eccentricity $e$, period $P$, and time of periastron $T_0$.  At each
grid point, the Thiele-Innes elements were determined by a linear fit
to the observational data using Singular Value Decomposition.  From
the Thiele-Innes elements, the semimajor axis $a$, the angle between
node and periastron $\omega$, the position angle of the line of nodes
$\Omega$, and the inclination $i$ were computed.

Since the orbit of LHS\,1070 B-C is already quite well-known
\citep{leinert01, seifahrt2008}, only a small range of parameter
values had to be scanned:
200 points within $0.018 \le e \le 0.028$, 200 points within
$17.21{\rm\,yr} \le P < 17.31{\rm\,yr}$, and initially 200 points for
$T_0$ distributed over one orbital period.
After the initial scan over $T_0$, the best estimate for $T_0$ was
improved by re-scanning a narrower range in $T_0$ centered on the
minimum found in the coarser scan.  This grid refinement was repeated
until the step size was less than one day.

%
%
%
\begin{table}
\caption{Parameters of the best orbital solution for the pair B-C.}
\label{OrbitBCtab}
\renewcommand{\arraystretch}{1.3}
\begin{center}
\begin{tabular}{lr@{}l}
\noalign{\vskip1pt\hrule\vskip1pt}
Orbital Element				& \multicolumn{2}{c}{Value} \\
\noalign{\vskip1pt\hrule\vskip1pt}
Date of periastron $T_0$			& $2454145.0$ & $\,^{+  1.4}_{  -0.8}$\\
						& (2007 Feb 13)\span\\
Period $P$ (years)				& $  17.24$ & $\,^{+ 0.01}_{ -0.01}$\\
Semi-major axis $a$ (mas)			& $  457.8$ & $\,^{+  0.3}_{  -0.4}$\\
Semi-major axis $a$ (AU)			& $   3.53$ & $\,^{+ 0.07}_{ -0.07}$\\
Eccentricity $e$				& $ 0.0227$ & $\,^{+0.0004}_{-0.0003}$\\
Argument of periastron $\omega$ ($^\circ$)	& $ 217.01$ & $\,^{+ 0.04}_{ -0.08}$\\
P.A. of ascending node $\Omega$ ($^\circ$)	& $  14.65$ & $\,^{+ 0.04}_{ -0.06}$\strut\\
Inclination $i$ ($^\circ$)			& $  61.82$ & $\,^{+ 0.04}_{ -0.04}$\\
{System mass $M_S$ ($\rm mas^3/year^2$)}	& $(3.23\pm{}$& $0.01)\cdot10^5$\\
System mass $M_B+M_C$ ($M_\odot$)		& $ 0.149$ & ${}\pm0.009$\\
Mean absolute difference\\
	\qquad observed -- predicted (mas)	& 4.9	\\
reduced $\chi^2$				& $    3.5$\\
\noalign{\vskip1pt\hrule\vskip1pt}
\end{tabular}
\end{center}
\end{table}

\begin{figure}[t]
\centerline{\includegraphics[width=\hsize]{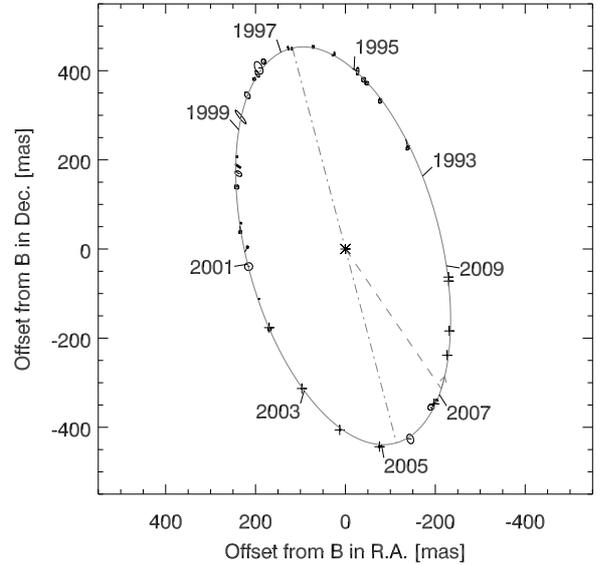}}
\caption{The orbit of component C around component B.
  The observed positions are marked by their error ellipses and lines
  connecting the observed and calculated position at the time of the
  observations.  The observations with NACO are marked by
  crosses. Their errors are too small to be discernible.
  The dash-dotted line indicates the line of nodes, the dashed line
  the periastron, and the arrow shows the direction of the orbital
  motion.}
\label{orbitBCpic}
\end{figure}

We improved the results of the grid-search with a Levenberg-Marquardt
$\chi^2$ minimization algorithm \citep{press92} that fits for all 7
parameters simultaneously.  The simple-minded approach would be to use
the orbital elements with the minimum $\chi^2$ found with the
grid-search.  However, initial test runs showed that the algorithm
does not converge on the global minimum.  For the same reason, we did
not use one of the previously published orbit solutions as starting
point \citep{leinert01,seifahrt2008}.  To make sure we find the
globally minimum $\chi^2$, we decided to use all orbits resulting from
the grid-search as starting points that had $\chi^2 < \chi^2_{\rm
  min}+9$.  The number 9 was chosen arbitrarily, to avoid starting
from obviously bad orbits.
The orbit with the globally minimum $\chi^2$ found in by the
Levenberg-Marquardt fit is shown in Fig.~\ref{orbitBCpic}, and its
elements are listed in Table~\ref{OrbitBCtab}.
To convert the semi-major axis from mas to AU, we used the distance
of $7.72\pm0.15\rm\,pc$ \citep{costa2005}.

{The reduced $\chi^2$ of 3.5 is higher than expected for a good
  fit, which indicates that we underestimate our astrometric errors.
  The mean absolute difference between observed and predicted
  positions is also larger than typical errors of our measurements.
  This might be caused by unknown errors in the calibration, as
  mentioned in section~\ref{ErrorSect}.  However, the residuals show
  no pattern indicating a systematic error, and the errors given in
  table~\ref{OrbitBCtab} are our best guess.  Therefore, we decided to
  accept the fit as it is.}

Errors of the orbital elements were determined by studying the
$\chi^2$ function around its minimum.  Since we are interested in the
confidence interval for each parameter taken separately, we have to
perturb one parameter (for example $T_0$) away from the minimum, and
optimize all the other parameters.  Any perturbation of a parameter
will of course lead to a larger $\chi^2$.  The range in $T_0$
within which $\chi^2(T_0) - \chi^2_{\rm min} < 1$ defines the 68\,\%
confidence interval for $T_0$.  This interval is usually not symmetric
around the $T_0$ of the best fit, therefore we list in
Table~\ref{OrbitBCtab} separate limits for positive and negative
perturbations.  It should be noted that these limits describe the
parameter range that contain 68\,\% of the probability distribution,
which is equivalent to the commonly used $1\sigma$-errors.  However,
the errors are not normally distributed, therefore a $2\sigma$
interval will {\em not\/} contain 95\,\% of the probability
distribution.

{Estimating the error of the mass required a special procedure.
  The mass itself is computed using Kepler's third law ($M=a^3/P^2$).
  The semi-major axis $a$ and the period $P$ are usually strongly
  correlated.  To obtain a realistic estimate for the mass
  error, we did {\em not\/} use the naive way of error propagation.
  Instead, we considered a set of orbital elements where the
  semi-major axis was replaced by the mass.  This is possible because
  Kepler's third law gives an unambiguous relation between the two
  sets of elements.  With the mass being one of the orbital elements,
  we can treat it as one of the independent fit parameters and
  determine its error with the method described in the previous
  paragraph.
}

With an orbit derived from astrometric measurements, there always
remains the ambiguity which of the two nodes is the ascending node
(defined as the node where the companion is receding from the
observer).  Fortunately, \citet{seifahrt2008} measured the relative
radial velocities of LHS\,1070 B and C on 2006 October 9.
They found it to be negative, i.e.\ LHS\,1070 C was approaching
us\footnote{It is not entirely clear whether \citet{seifahrt2008} give
the velocity of C minus B, but they report $\Omega = 14.5^\circ$,
which is consistent with our interpretation}.
Therefore, the position angle of the ascending node is $14.65^\circ$
(cf.\ Fig.~\ref{orbitBCpic}), and the position angle of the descending
node is $194.65^\circ$.
{Our orbit model predicts a relative radial velocity of
  $-4.8\pm0.1\rm\,km/s$ on 2006 Oct 9, in agreement with the
  measurement of \cite{seifahrt2008}.}


\section{The orbit of LHS 1070 BC around A}
\label{orbitABCsect}

The orbit of components B and C around each other allows us to
determine only the combined mass of B and C.  To compute the
individual masses, we need to know the mass ratio $q$, which can be
computed if the position of the center of mass (CM) of B and C is
known.  Unfortunately, we cannot observe the CM directly.  However, we
know that the CM of B and C is in orbit around component A\footnote{%
  We did not detect the companion D reported by \citet{henry1999}.
  Therefore, by ``A'' we mean the suspected close binary composed of A
  and D (if it exists) in the nomenclature of \citet{henry1999}.},
and that B and C are in orbit around their CM.
The CM is always on the line between B and C, and its distance from B
is the constant fraction $q/(1+q)$ of the separation of B and C.

%
%
%
\begin{table}
\caption{Parameters of the best solution for the orbit of BC around A.}
\label{OrbitABCtab}
\renewcommand{\arraystretch}{1.3}
\begin{center}
\begin{tabular}{lr@{}l}
\noalign{\vskip1pt\hrule\vskip1pt}
Orbital Element					 & \multicolumn{2}{c}{Value} \\
\noalign{\vskip1pt\hrule\vskip1pt}
Date of periastron $T_0$			 & $2459727.9$ & $\,^{+  3.6}_{  -3.1}$\\ 
						 & (2022 May 28)\span\\
Period $P$ (years)				 & $   44.4$ & $\,^{+11.9}_{ -2.4}$\\ 
Semi-major axis $a$ (mas)			 & $ 1111.6$ & $\,^{+  0.8}_{  -1.1}$\\ 
Semi-major axis $a$ (AU)			 & $   8.58$ & $\,^{+ 0.17}_{ -0.17}$\\
Eccentricity $e$				 & $ 0.5200$ & $\,^{+0.0020}_{-0.0012}$\\ 
Argument of periastron $\omega$ ($^\circ$)	 & $ 147.61$ & $\,^{+ 7.94}_{ -0.55}$\\ 
P.A. of ascending node $\Omega$ ($^\circ$)	 & $  26.80$ & $\,^{+ 0.08}_{ -0.08}$\\ 
Inclination $i$ ($^\circ$)			 & $  54.75$ & $\,^{+ 0.90}_{ -0.10}$\\ 
{System mass $M_A+M_B+M_C$ ($\rm mas^3/yr^2$)}&$(6.98\pm{}$&$0.02)\cdot10^5$ \\
System mass $M_A+M_B+M_C$ ($M_\odot$)		& $0.321$ & $\,^{+0.020}_{-0.028}$\\
Mass ratio $M_{\rm C}/M_{\rm B}$			& $0.92$  & ${}\pm 0.01$\\
Mean absolute difference\\
	\qquad observed -- predicted (mas)	& 7.9	\\
reduced $\chi^2$				& $    2.6$\\
\noalign{\vskip1pt\hrule\vskip1pt}
Mass of A $M_{\rm A}$ ($M_\odot$)		& $0.172$ & ${}\pm 0.010$\\
Mass of B $M_{\rm B}$ ($M_\odot$)		& $0.077$ & ${}\pm 0.005$\\
Mass of C $M_{\rm C}$ ($M_\odot$)		& $0.071$ & ${}\pm 0.004$\\
\noalign{\vskip1pt\hrule\vskip1pt}
\end{tabular}
\end{center}
\end{table}



We follow the method that was used by \citet{koehler2008} to derive
masses in the triple system T~Tauri.
The position of the CM of B and C is described in two ways:
First, it is on a Keplerian orbit around A, which is described by 7
orbital elements.
Second, the position of the CM can be computed from the observed
positions of B and C, and the mass ratio (which is treated as a free
parameter).
Standard error propagation is used to obtain an error estimate for
this position.  To compute $\chi^2$, we compare the position of the CM
from the orbit around A with the positions derived from the
observations.
Our model has therefore 8 free parameters, the 7 elements which
describe the orbit of the CM of B+C around A, and the parameter
$f=q/(1+q)$.  The parameter $f$ is often called fractional mass
\citep{heintz78}, since it is the secondary star's fraction of the
total mass in a binary.  It is useful in our case because it also
describes the fractional offset of the CM from B, i.e.\ the
separation between B and the CM divided by the separation between B
and C.  For a grid-search, $f$ is better suited than $q$,
because $f$ is confined to the range 0 to 1, while $q$ is a number
between 0 and infinity.


\begin{figure}[t]
\centerline{\includegraphics[width=\hsize]{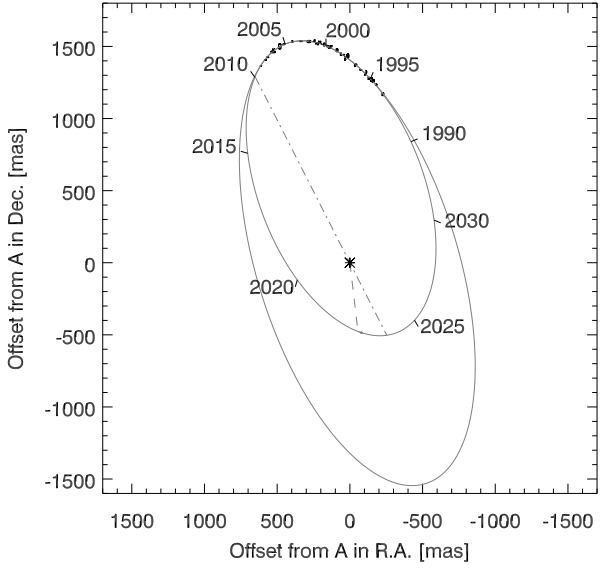}}
\caption{The orbit of the center of mass of components B and C around
  component A.
  The observed positions are marked by their error ellipses and
  lines connecting the observed and calculated position at the
  time of the observations.  Note that these positions are computed
  from the observed positions of components B and C, and the mass
  ratio $q$, which is a free parameter of the model fit.
  The dash-dotted line indicates the line of nodes, the dashed line
  the periastron, and the arrow shows the direction of the orbital
  motion.
  The two orbits shown are the model with the globally minimum
  $\chi^2$, and an almost circular orbit with a longer period of about
  80\,years.}
\label{orbitABCpic}
\end{figure}

\begin{figure}[t]
\centerline{\includegraphics[width=\hsize]{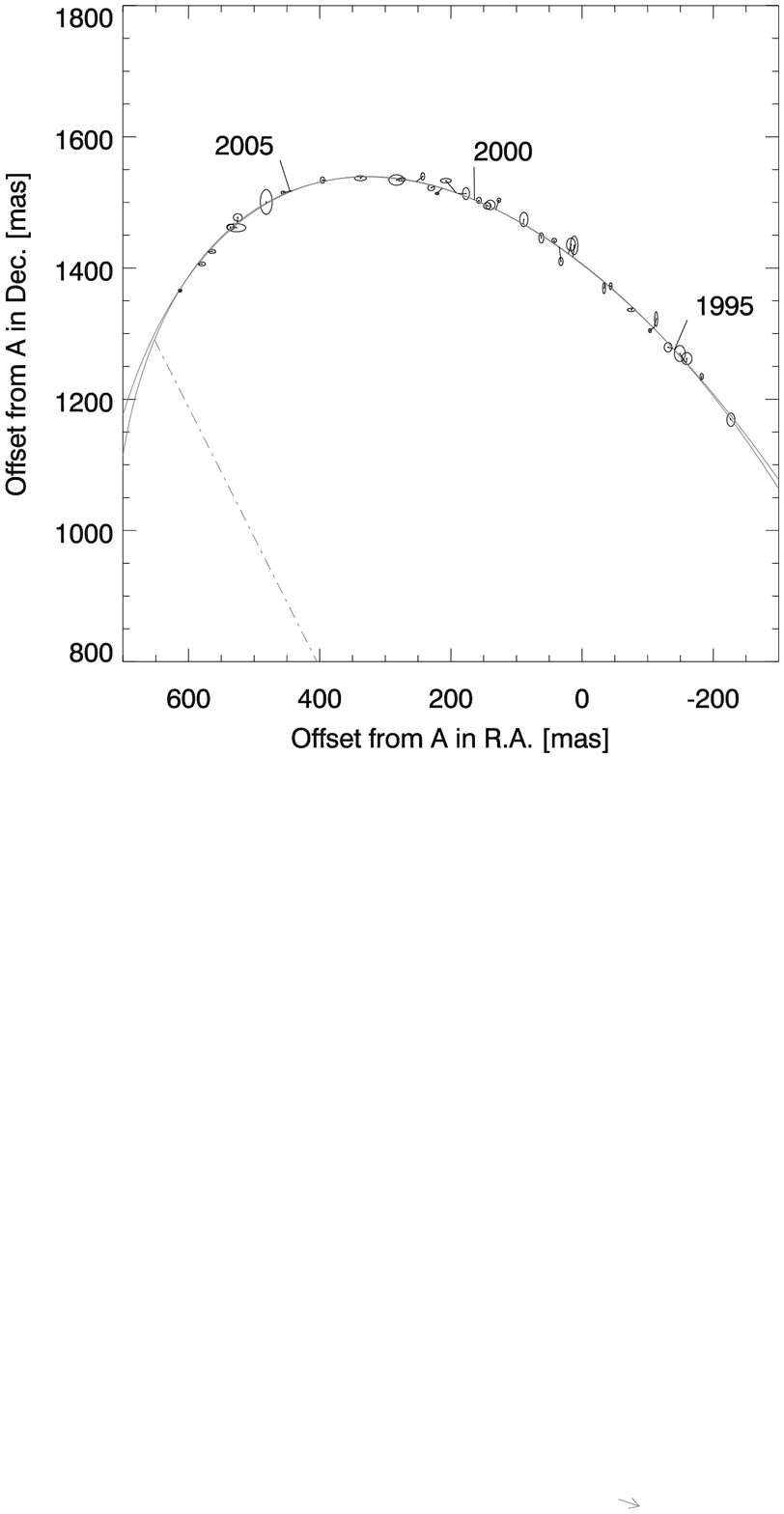}}
\caption{Enlarged section of the orbits in Fig.~\ref{orbitABCpic},
  showing the part covered by observations.}
\label{orbitABCxpic}
\end{figure}


The fitting procedure is similar to that used for the orbit of B-C,
except that the grid-search is carried out in 4 dimensions:
eccentricity $e$, period $P$, time of periastron $T_0$, and the
fractional mass~$f$.
Singular Value Decomposition was used to fit the Thiele-Innes
constants, which give the remaining orbital elements.
It is worth noting that the orbital elements in this fit describe the
orbit of the A-BC binary, only the fractional mass $f$ refers to the
pair B+C.

For the four dimensional grid search, we used 100 points each for $f$,
$e$, and $P$.  The grid ranged from 0 to 0.99 in $f$, 0 to 0.99 in
$e$, and 30 to 300 years in $P$.  The grid in $T_0$ started with 100
points distributed uniformly over one orbital period.  Similar to the
fit for the orbit of B-C, the grid in $T_0$ was refined until the grid
spacing was less than one day.

The results of the grid-search were improved with a
Levenberg-Marquardt $\chi^2$ minimization algorithm \citep{press92}.
As starting points, we used all orbits resulting from the grid-search
that had $\chi^2 < \chi^2_{\rm min}+9$.
The orbit with the globally minimum $\chi^2$ found by the
Levenberg-Marquardt fit is shown in Figs.~\ref{orbitABCpic} and
\ref{orbitABCxpic}.
Table~\ref{OrbitABCtab} lists its elements, as well as the mass ratio
$q$, and the individual masses of all three components.
{As with the inner orbit of B and C, $\chi^2$ is somewhat larger then
 expected for a good fit.}


\begin{figure}[ht]
\centerline{\includegraphics[width=\hsize]{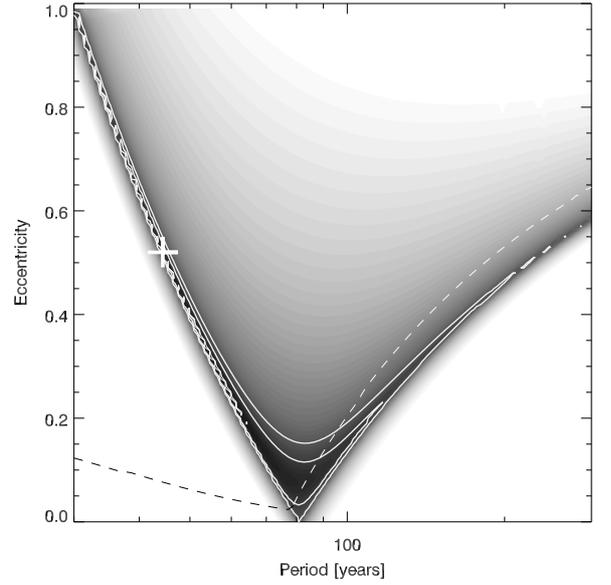}}
\caption{$\chi^2$ as function of $P$ and $e$ for the orbit of BC
  around A.
  The cross at $P=44.4\rm\,yr$, $e=0.52$ marks the minimum.
  The contour lines encircle the 68\,\% confidence region (which is
  a series of mostly unconnected patches), the 95\,\%, and the
  99.7\,\% confidence region (corresponding to $1\sigma$, $2\sigma$,
  and $3\sigma$ in the case of normally distributed errors).
  The dashed line marks the stability limit for the system, only
  orbits below this line are long-term stable (see
  section~\ref{StabilitySect}).
}
\label{chiABCpic}
\end{figure}


Errors for the parameters were again estimated by analyzing the
$\chi^2$ function around the minimum.  Since only a relatively small
fraction of the orbit has been observed so far, the uncertainties for
the orbital elements are much larger than for the orbit of B and C
around each other.  Figure~\ref{chiABCpic} shows $\chi^2$ as function of
period and eccentricity.  Orbits with periods of more than 100\,years
are within the 95\,\% confidence region, and periods of more than
200\,years are in the 99.7\,\% confidence region.
{The error of the system mass was derived in the same way as with
  the orbit of LHS\,1070\,B and C around each other, i.e.\ by treating
  $M$ as an independent fit parameter.}


The ambiguity of the ascending node is resolved with the help of
\cite{seifahrt2008}. They measured in 2006 a positive relative radial
velocity between the barycentre of LHS\,1070 B and C and component A.
Therefore, the binary BC was receding from the observer, and the
position angle of the ascending node must be $26.8^\circ$ (cf.\
Fig.~\ref{orbitABCpic}).
{Our orbit model predicts a relative radial velocity of
  $3.1\pm0.1\rm\,km/s$ on 2006 Oct 9, in agreement with the
  measurement of \cite{seifahrt2008}.}

Finally, the mass of component A can be computed from the system mass
of the triple and the mass of the binary B+C (derived in section
\ref{orbitBCsect}).  The largest
contribution to the error of both masses is the uncertainty in the
distance to the system.  Since this contribution to the errors of
$M_{\rm ABC}$ and $M_{\rm BC}$ is correlated, it would not be correct
to add the errors in quadrature.  Instead, we computed masses from the
semi-major axes in mas (which results in the unusual mass unit of $\rm
mas^3/years^2$), subtracted $M_{\rm BC}$ from $M_{\rm ABC}$, and
converted the resulting $M_{\rm A}$ into solar masses by multiplying
it with the distance cubed.  This way, the error of the distance
enters the calculation only once, resulting in the correct estimate
for the error of $M_{\rm A}$.


\section{Discussion}

\subsection{Is the LHS\,1070 system stable?}
\label{StabilitySect}

\begin{figure}[ht]
\centerline{\includegraphics[width=\hsize]{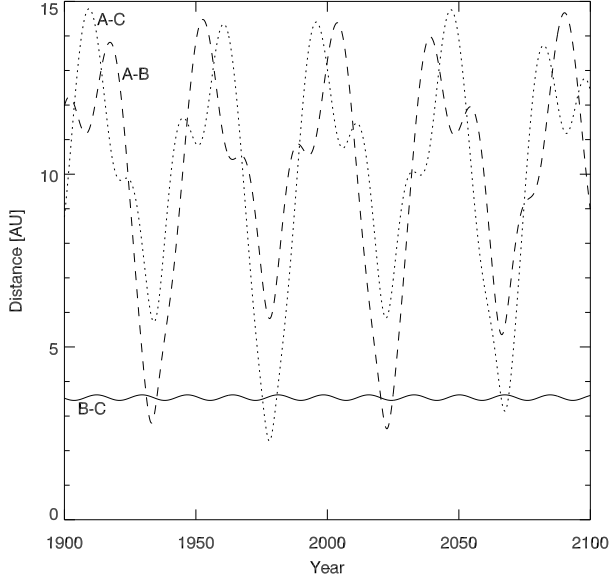}}
\caption{Distances between the three stars in the system as function
  of time.  Here we assume that the stars stay on the Keplerian orbits
  derived in sections~\ref{orbitBCsect} and \ref{orbitABCsect}, i.e.\ we
  do not simulate the three-body problem posed by the triple system.}
\label{distABCpic}
\end{figure}

Our best fit for the orbit of the binary BC around A is only a factor
2.5 larger than the orbit of BC itself.  The outer orbit has also a
rather large eccentricity.  As a result, the distance from B or C to A
becomes smaller than the distance to its binary companion during the
periastron passage of the outer orbit (Fig.~\ref{distABCpic}).  This
raises the question whether a triple system like this can be stable
over timescales comparable to its age.


\begin{figure}[t]
\centerline{\includegraphics[width=\hsize]{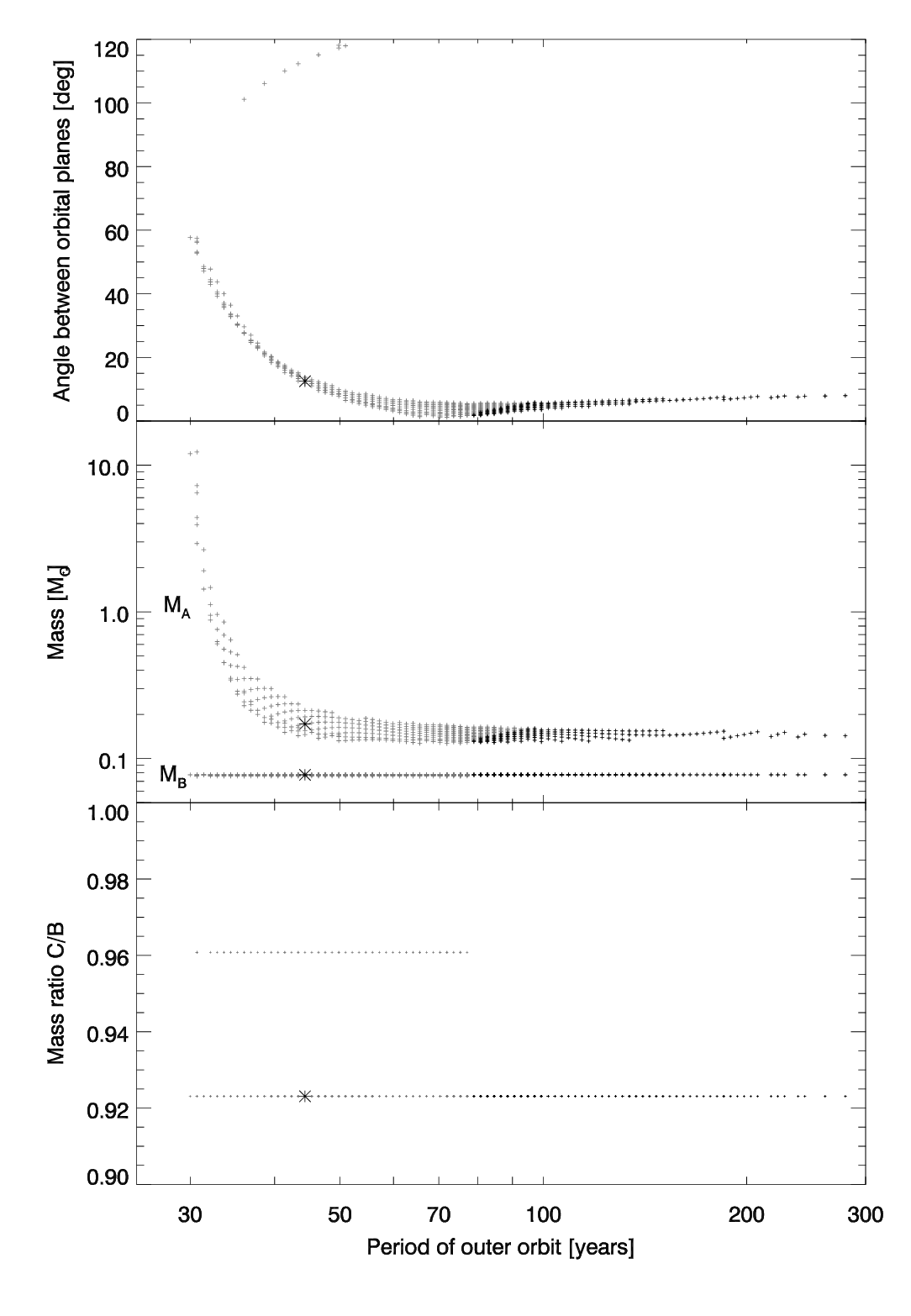}}
\caption{Results of the fits for the outer orbit as function of their
  period.  Shown are all model orbits within the 99\,\% confidence
  region ($3\sigma$).
  Unstable orbits \citep[according to the criterion by][]{donnmik1995}
  are indicated by gray symbols, stable orbits are black.
  The orbit with the minimal $\chi^2$ is marked by an asterisk.
}
\label{massABCpic}
\end{figure}


\citet{donnmik1995} presented a criterion for the long-term stability
of coplanar triple systems:
The periastron distance $q_2 = a_2(1-e_2)$ of the outer orbit should
be at least a factor of 3.3 larger than the semi-major axis $a_1$ of
the inner orbit.
In non-coplanar systems, increasing the inclination decreases the
region of stability, i.e.\ the outer orbit has to be even larger to be
stable \citep{donnison2009}.
In Fig.~\ref{chiABCpic}, the limit for coplanar orbits
($q_2 > 3.3\,a_1$) is indicated by the dashed line, only orbits below
the line are stable.
The orbit with the minimum $\chi^2$ is clearly in the unstable regime.
However, the $2\sigma$ and $3\sigma$ confidence regions extend well
into the regime of stable orbits.
It is therefore possible that the true orbit has a longer period of at
least 80\,years (cf.\ Fig.~\ref{orbitABCpic}).

How does this uncertainty about the orbit influence our conclusions?
Figure~\ref{massABCpic} shows the parameters of the outer orbit that
we are most interested in.
Shown are all orbits within the 99\,\% confidence region in our grid
of possible solutions.
Unstable orbits \citep[according to the criterion by][]{donnmik1995}
are indicated by gray symbols, stable orbits are black.
The top panel shows the angle between the planes of the inner and
outer orbits.  This angle is $12.5^\circ$ for the orbit with the
minimum $\chi^2$, and less than $10^\circ$ for all stable orbits.
We find a few orbital solutions with angles larger than $100^\circ$,
but all these orbits are unstable.  We therefore conclude that the
inner and outer orbit are almost coplanar.  Therefore, the stability
criterion by \citet{donnmik1995} is sufficient in our case, although
it does not apply to inclined orbits.

The second panel of Fig.~\ref{massABCpic} shows the masses of
LHS\,1070\,A and B.  Orbits with short periods result in
unrealistically high masses for A, which also indicates that the
true period is longer.  The mass of component B lies between $0.076$
and $0.077\rm\,M_\odot$, almost independent of the orbital period.
The mass of component C was not plotted, since it can easily be
computed as $0.149{\rm\,M_\odot} - M_B$.

Finally, the bottom panel of Fig.~\ref{massABCpic} shows the mass
ratio between LHS\,1070\,C and B.  All stable orbits result in a mass
ratio of 0.923, which is the same as the mass ratio of the orbit with
the minimum $\chi^2$.


We conclude that it is possible to find stable orbital solutions that
are compatible with the astrometric measurements.  In these stable
configurations, the orbit of B and C around each other, and the orbit
of B+C around A are almost coplanar, with angles of less than
$10^\circ$.  The mass of component A lies between $0.13$ and
$0.16\rm\,M_\odot$ for stable orbits.

{As a final note on the stability of the system, we point out
  that the stability criterion used here does not take into account
  resonances.  For example, a period of the outer orbit of 52 years
  would be in a 3:1 resonance with the inner orbit.  This orbit would
  be within the $1\sigma$ region around the orbit model with minimum
  $\chi^2$.  However, the uncertainties of our orbit fit are too large
  to provide a reasonable starting point for a search for stable
  resonant configurations.  Furthermore, it would not significantly
  change our conclusions about coplanarity and the masses of the three
  components if the period of the outer orbit was between about 50 and
  80\,years.}


\subsection{Comparison with theoretical models}

\begin{figure}[ht]
\centerline{\includegraphics[angle=90,width=\hsize]{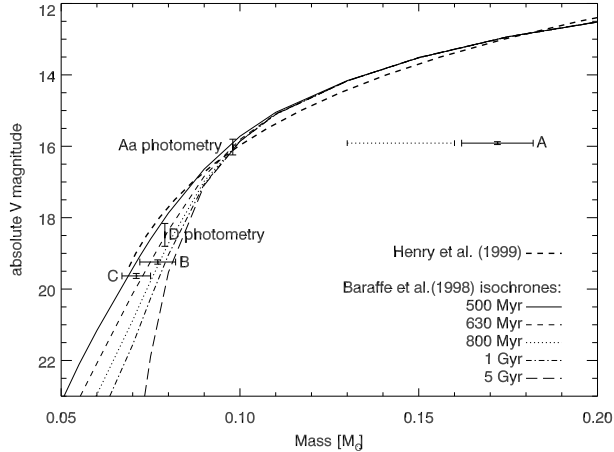}}
\caption{Mass-luminosity diagram with the components of LHS\,1070,
  the models of \citet{BCAH98}, {and the empirical
  Mass-Luminosity-Relation by \citet{henry1999}.
  Shown are the theoretical} isochrones for ages between 500\,Myr and
  5\,Gyr.
  The dotted horizontal line indicates the mass range for A resulting
  from stable orbit configurations.  The points labeled ``Aa
  photometry'' and ''D photometry'' indicate the magnitudes if A is
  split into two components \citep{henry1999}, with the masses
  adjusted to put the points onto the 630-Myr isochrone.}
\label{BCAHpic}
\end{figure}

The goal of dynamical mass determinations is to test theoretical
predictions for the mass and luminosity of the stars.
Figure~\ref{BCAHpic} shows a mass-luminosity-diagram with the
isochrones of \citet{BCAH98} and our results for the components of
LHS\,1070 (using the V-magnitudes given by \citealt{leinert2000}).
{Additionally, it shows the empirical Mass-Luminosity-Relation by
\citet{henry1999}, which is similar to the theoretical
500-Myr-isochrone. 
}
Within the error bars, both LHS\,1070\,B and C are compatible with an
age between 500 and 800\,Myr.

On the other hand, LHS\,1070\,A is either $3^{\rm mag}$ too faint or
$0.07\rm\,M_\odot$ too heavy compared to the theoretical models
{and the empirical relation}.
This is also true if we take into account that stable orbit
configurations result in lower masses (indicated by the dotted line in
Fig.~\ref{BCAHpic}).
In the K-band, LHS\,1070\,A is about $1^{\rm mag}$ too faint compared
to the model isochrones.  This excludes the possibility that the
under-luminosity is due to extinction by circumstellar or foreground
material (as unlikely as it may be for a star of this age and
distance), since then the effect in the infrared should be smaller,
only about $0.3^{\rm mag}$.

The simplest explanation would be that A is indeed a binary as
reported by \citet{henry1999}.  In the following, we will call the
components Aa and D, and continue to use A for the (unresolved) binary
system.
We computed the absolute V-magnitudes of Aa and D from the combined
apparent magnitude $V=15.35^{\rm mag}$ \citep{leinert2000} and the
magnitude difference $\Delta V=2.46^{\rm mag}$ \citep{henry1999},
using the distance of 7.72\,pc \citep{costa2005}.  By finding the
intersection with the 630-Myr isochrone from the models of
\citet{BCAH98}, we can estimate the mass of Aa and D to be
$0.10\rm\,M_\odot$ and $0.08\rm\,M_\odot$, resp.  The combined mass
of A resulting from photometry is therefore about $0.18\rm\,M_\odot$,
in reasonable agreement with the dynamical mass resulting from the
orbit determinations.


In an attempt to find a possible close companion to LHS\,1070\,A, we
analyzed all available high-resolution data of the system.
The resolution required to detect component D can only be reached with
a 8\,m-class telescope, preferentially at short wavelengths,
e.g. J-band.  This means we have to rely on NACO data.  There are
two datasets taken with NACO in the J-band on June 27.\ and
December 12., 2003, and two datasets in the narrow-band filter NB\_2.17
in a mode that allows speckle post-processing of the data.  In all four
datasets, we can exclude an companion with a flux ratio brighter than
0.2, i.e.\ a magnitude difference of $1.75^{\rm mag}$.  The resolution
of the four observations is 30 -- 40\,mas.  With the masses of Aa and
D derived above, the magnitude difference should be about $0.8^{\rm
  mag}$.  Assuming a semi-major axis of 50\,mas and a system mass of
$0.17\rm\,M_\odot$ results in an orbital period of about
$0.6\rm\,years$.  It is possible that all our observations were
carried out at times when the companion was too close to the primary
to be visible.  However, given the large number of attempt to
detect LHS\,1070\,D this appears unlikely.


{We also looked into the possibility to detect a periodic
  astrometric signal of LHS\,1070\,D in the residuals of the A-BC
  separation.  If LHS\,1070\,A is an unresolved binary, then an
  astrometric shift of its center of light with the period of its
  orbit can occur.  The magnitude of this shift depends on
  the position of the center of mass and the position of the center of
  light, hence the mass ratio and the flux ratio.  The mass ratio of
  our suspected Aa-D binary is 0.8, while the flux ratio computed from
  the models of \citet{BCAH98} is 0.48.  The expected motion of the
  center of light is therefore about 0.12 times the size of the orbit
  of D around Aa (6\,mas for an orbit of 50\,mas).
  Unfortunately, our sampling is not very dense, with 38 observations
  in 15.25\,years.  The Nyquist frequency $\nu_N = N/(2T)$
  corresponds to a period of 0.8\,years.  Since our sampling is rather
  uneven, the Nyquist frequency is not a sharp limit for the
  detectability of a periodic signal.  Nonetheless, due to the sparse
  sampling, it is unlikely that we can detect periods much shorter
  than 0.8\,years.  In fact, neither classical nor generalized
  Lomb-Scargle-Periodograms \citep{scargle1982,zechmeister2009} show
  any significant peaks.
  Even adding an artificial signal with a period of 0.6\,years and a
  large amplitude of 100\,mas does not result in a significant
  detection, although peaks at 1/2 and 1/4 of the frequency appear.
  In summary, we conclude that an astrometric detection of
  LHS\,1070\,D is not possible with the available data.
}

Another possible explanation could be that LHS\,1070\,A is a binary too
close to be resolved by the observations available so far.  This
hypothetical companion would not be identical with component D.
However, at the moment the most likely explanation might be that we
overestimate the system mass.  This would not be too surprising, given
the small observational coverage of the orbit.


\section{Summary and Conclusions}

We present new relative positions of LHS\,1070\,A, B, and C, collected
in the years 2000 to 2008.  They were used to derive an improved model
for the orbit of B and C around each other, and an estimate for the
orbit of B and C around A.  The orbit of B and C is well-determined by
now, with an orbital period of $17.24\pm0.01$\,years and a system mass
of $0.149\pm0.009\rm\,M_\odot$ (cf.\ Table \ref{OrbitBCtab}).

The observations of LHS\,1070\,B relative to A span only a range in
position angle of $36^\circ$.  The orbit of B and C around A derived
from these observations is therefore not so well-constrained.  The
orbital solution minimizing $\chi^2$ has a rather short period of
about 44\,years, but it results in an unstable configuration of the
triple system.  It is more likely that the true orbital period is in
the range 80 -- 200 years.  Orbital solutions in this range of periods
are still within the 95\,\% and 99.7\,\% confidence region for the
fit to the astrometric observations (which corresponds to $2\sigma$
and $3\sigma$ for the case of normally distributed measurement
errors).  Despite this uncertainly, we can constrain the system mass
$M_A+M_B+M_C$ to 0.28 to 0.31 M$_\odot$ if we accept only stable orbit
configurations.

The outer orbit also yields the mass ratio of B and C, which is quite
well-constrained to $0.92\pm0.01$.  Taken together, these results
yield individual masses of $M_A = 0.13\ldots0.16\rm\,M_\odot$,
$M_B = 0.077\pm0.005\rm\,M_\odot$ and
$M_C = 0.071\pm0.004\rm\,M_\odot$.

Placing the three stars in a mass-luminosity diagram and comparing
with theoretical isochrones shows that B and C are coeval within the
measurement errors, with an age between 500 and 800\,Myr.
On the other hand, LHS\,1070\,A appears to be too faint for its mass,
or too massive for its luminosity.  One possible explanation could be
that LHS\,1070\,A is itself a binary.  There has been one report of
the discovery of a close companion \citep{henry1999}, but this
discovery could not be confirmed, despite the large number of
observations collected.  The detection might have been caused by a
glitch in the data (T.~Henry, priv.\ comm.).
To confirm or disprove the binarity of LHS\,1070\,A, observations
sensitive to companions at very small separations would be helpful,
either spectroscopic or interferometric.
For the time being, the most likely explanation might be that our
orbit fit overestimates the system mass.

Our results for the masses of the three stars show that LHS\,1070\,C
is almost certainly a brown dwarf.
LHS\,1070\,B is very close to the hydrogen-burning mass-limit,
possibly also a brown dwarf.
While LHS\,1070\,A is clearly above the hydrogen-burning limit, it is
still a very low mass star.
The very low mass triple system LHS\,1070 thus has the very
interesting property to contain one component above, one below, and
one at the hydrogen burning limit. This combination makes it an ideal
candidate for testing the change of atmosphere properties in this mass
regime.
Furthermore, it allows some conclusions about its formation.  One
theory for the formation of brown dwarfs is that they got ejected out
of a multiple system before they could accrete enough mass to start
hydrogen burning \citep{reipurth2001}.  It is difficult how a system
like LHS\,1070 can survive such an event.  Even if it managed to
remain bound, one would expect more eccentric and inclined orbits than
found in this work. It is more likely that LHS\,1070 formed in the
same way as normal stars, where a number of triple and higher-order
multiple systems are known.


\begin{acknowledgements}
  We thank the many observers and support astronomers who carried out
  the many separate observations necessary for this work.
  We are also grateful for the helpful comments by the editor.


\end{acknowledgements}

\bibliographystyle{bibtex/aa}
\bibliography{LHS1070}

\begin{thebibliography}{28}
\expandafter\ifx\csname natexlab\endcsname\relax\def\natexlab#1{#1}\fi

\bibitem[{{Almeida} {et~al.}(2011){Almeida}, {Jablonski}, \&
  {Martioli}}]{almeida2011}
{Almeida}, L.~A., {Jablonski}, F., \& {Martioli}, E. 2011, \aap, 525, A84

\bibitem[{{Baraffe} {et~al.}(1998){Baraffe}, {Chabrier}, {Allard}, \&
  {Hauschildt}}]{BCAH98}
{Baraffe}, I., {Chabrier}, G., {Allard}, F., \& {Hauschildt}, P.~H. 1998, \aap,
  337, 403

\bibitem[{{Basri} \& {Marcy}(1995)}]{basri1995}
{Basri}, G. \& {Marcy}, G.~W. 1995, \aj, 109, 762

\bibitem[{{Costa} {et~al.}(2005){Costa}, {M{\'e}ndez}, {Jao}, {Henry},
  {Subasavage}, {Brown}, {Ianna}, \& {Bartlett}}]{costa2005}
{Costa}, E., {M{\'e}ndez}, R.~A., {Jao}, W.-C., {et~al.} 2005, \aj, 130, 337

\bibitem[{Diolaiti {et~al.}(2000)Diolaiti, Bendinelli, Bonaccini, Close,
  Currie, \& Parmeggiani}]{Diolaiti00}
Diolaiti, E., Bendinelli, O., Bonaccini, D., {et~al.} 2000, \aaps, 147, 335

\bibitem[{{Donnison}(2009)}]{donnison2009}
{Donnison}, J.~R. 2009, \planss, 57, 771

\bibitem[{{Donnison} \& {Mikulskis}(1995)}]{donnmik1995}
{Donnison}, J.~R. \& {Mikulskis}, D.~F. 1995, \mnras, 272, 1

\bibitem[{Heintz(1978)}]{heintz78}
Heintz, W.~D. 1978, Double Stars, Geophysics and Astrophysics Monographs No.~15
  (Dordrecht, Holland: D.\ Reidel Publishing Company)

\bibitem[{{Henry} {et~al.}(1999){Henry}, {Franz}, {Wasserman}, {Benedict},
  {Shelus}, {Ianna}, {Kirkpatrick}, \& {McCarthy}}]{henry1999}
{Henry}, T.~J., {Franz}, O.~G., {Wasserman}, L.~H., {et~al.} 1999, \apj, 512,
  864

\bibitem[{{Hofmann} {et~al.}(1992){Hofmann}, {Blietz}, {Duhoux}, {Eckart},
  {Krabbe}, \& {Rotaciuc}}]{SHARP}
{Hofmann}, R., {Blietz}, M., {Duhoux}, P., {et~al.} 1992, in Progress in
  Telescope and Instrumentation Technologies, ESO Conference and Workshop
  Proceedings No. 42, ed. M.-H. {Ulrich} (ESO Garching), 617

\bibitem[{{Hofmann} {et~al.}(1995){Hofmann}, {Brandl}, {Eckart}, {Eisenhauer},
  \& {Tacconi-Garman}}]{SHARPII}
{Hofmann}, R., {Brandl}, B., {Eckart}, A., {Eisenhauer}, F., \&
  {Tacconi-Garman}, L.~E. 1995, in Society of Photo-Optical Instrumentation
  Engineers (SPIE) Conference Series, Vol. 2475, Society of Photo-Optical
  Instrumentation Engineers (SPIE) Conference Series, ed. {A.~M.~Fowler},
  192--202

\bibitem[{K{\"o}hler {et~al.}(2000)K{\"o}hler, Kunkel, Leinert, \&
  Zinnecker}]{koehler2000}
K{\"o}hler, R., Kunkel, M., Leinert, C., \& Zinnecker, H. 2000, \aap, 356, 541

\bibitem[{{K{\"o}hler} {et~al.}(2008){K{\"o}hler}, {Ratzka}, {Herbst}, \&
  {Kasper}}]{koehler2008}
{K{\"o}hler}, R., {Ratzka}, T., {Herbst}, T.~M., \& {Kasper}, M. 2008, \aap,
  482, 929

\bibitem[{{Leinert} {et~al.}(2000){Leinert}, {Allard}, {Richichi}, \&
  {Hauschildt}}]{leinert2000}
{Leinert}, C., {Allard}, F., {Richichi}, A., \& {Hauschildt}, P.~H. 2000, \aap,
  353, 691

\bibitem[{{Leinert} {et~al.}(2001){Leinert}, {Jahrei{\ss}}, {Woitas}, {Zucker},
  {Mazeh}, {Eckart}, \& {K{\"o}hler}}]{leinert01}
{Leinert}, C., {Jahrei{\ss}}, H., {Woitas}, J., {et~al.} 2001, \aap, 367, 183

\bibitem[{{Leinert} {et~al.}(1994){Leinert}, {Weitzel}, {Richichi}, {Eckart},
  \& {Tacconi-Garman}}]{leinert1994}
{Leinert}, C., {Weitzel}, N., {Richichi}, A., {Eckart}, A., \&
  {Tacconi-Garman}, L.~E. 1994, \aap, 291, L47

\bibitem[{{Lenzen} {et~al.}(1998){Lenzen}, {Bizenberger}, {Salm}, \&
  {Storz}}]{lenzen98}
{Lenzen}, R., {Bizenberger}, P., {Salm}, N., \& {Storz}, C. 1998, in Society of
  Photo-Optical Instrumentation Engineers (SPIE) Conference Series, Vol. 3354,
  Society of Photo-Optical Instrumentation Engineers (SPIE) Conference Series,
  ed. {A.~M.~Fowler}, 493--499

\bibitem[{Lenzen {et~al.}(2003)Lenzen, Hartung, Brandner, Finger, Hubin,
  Lacombe, Lagrange, Lehnert, Moorwood, \& Mouillet}]{lenzen03}
Lenzen, R., Hartung, M., Brandner, W., {et~al.} 2003, in Instrument Design and
  Performance for Optical/Infrared Ground-based Telescopes, ed. M.~Iye \&
  A.~F.~M. Moorwood, SPIE Proceedings No. 4841, 944--952

\bibitem[{McCaughrean \& Stauffer(1994)}]{mccaughrean94}
McCaughrean, M.~J. \& Stauffer, J.~R. 1994, \aj, 108, 1382

\bibitem[{Press {et~al.}(1992)Press, Teukolsky, Vetterling, \&
  Flannery}]{press92}
Press, W.~H., Teukolsky, S.~A., Vetterling, W.~T., \& Flannery, B.~P. 1992,
  Numerical Recipes in C, 2nd edn. (Cambridge, UK: Cambridge University Press)

\bibitem[{{Reiners} {et~al.}(2007){Reiners}, {Seifahrt}, {K{\"a}ufl},
  {Siebenmorgen}, \& {Smette}}]{reiners2007}
{Reiners}, A., {Seifahrt}, A., {K{\"a}ufl}, H.~U., {Siebenmorgen}, R., \&
  {Smette}, A. 2007, \aap, 471, L5

\bibitem[{{Reipurth} \& {Clarke}(2001)}]{reipurth2001}
{Reipurth}, B. \& {Clarke}, C. 2001, \aj, 122, 432

\bibitem[{{Rousset} \& {Beuzit}(1999)}]{ADONIS}
{Rousset}, G. \& {Beuzit}, J.-L. 1999, {The COME-ON/ADONIS systems}, ed.
  {Roddier, F.}, 171

\bibitem[{Rousset {et~al.}(2003)Rousset, Lacombe, Puget, Hubin, Gendron, Fusco,
  Arsenault, Charton, Feautrier, Gigan, Kern, Lagrange, Madec, Mouillet,
  Rabaud, Rabou, Stadler, \& Zins}]{rousset03}
Rousset, G., Lacombe, F., Puget, P., {et~al.} 2003, in Adaptive Optical System
  Technologies II, ed. P.~L. Wizinowich \& D.~Bonaccini, SPIE Proceedings No.
  4839, 140--149

\bibitem[{{Scargle}(1982)}]{scargle1982}
{Scargle}, J.~D. 1982, \apj, 263, 835

\bibitem[{{Seifahrt} {et~al.}(2008){Seifahrt}, {R{\"o}ll}, {Neuh{\"a}user},
  {Reiners}, {Kerber}, {K{\"a}ufl}, {Siebenmorgen}, \& {Smette}}]{seifahrt2008}
{Seifahrt}, A., {R{\"o}ll}, T., {Neuh{\"a}user}, R., {et~al.} 2008, \aap, 484,
  429

\bibitem[{{van Altena} {et~al.}(1995){van Altena}, {Lee}, \&
  {Hoffleit}}]{altena1995}
{van Altena}, W.~F., {Lee}, J.~T., \& {Hoffleit}, E.~D. 1995, {The general
  catalogue of trigonometric [stellar] parallaxes}, ed. {van Altena, W.~F.,
  Lee, J.~T., \& Hoffleit, E.~D.}

\bibitem[{{Zechmeister} \& {K{\"u}rster}(2009)}]{zechmeister2009}
{Zechmeister}, M. \& {K{\"u}rster}, M. 2009, \aap, 496, 577

\end{thebibliography}

\end{document}